\RequirePackage{fix-cm}
\documentclass{svjour3}                     
\smartqed  
\usepackage{graphicx}
\usepackage{natbib}
\usepackage{url}
\usepackage[breaklinks,colorlinks=true,linkcolor=black,anchorcolor=black,citecolor=black,urlcolor=black]{hyperref}

\begin{document}

\title{Fuck the Algorithm \thanks{* Thanks to Phoebe Friesen, Leif Hancox-Li, Kathryn Lindeman, and Luke Stark for detailed comments on an early draft, and to the students in my 2021 Topics in Philosophy of Science: 3rd Wave AI course at Queen's University (especially Robyn Elliott and Trevell Hamilton-Frederick) for helpful discussion of a later draft.}
}
\subtitle{Conceptual Issues in Algorithmic Bias}


\author{Catherine Stinson}
\institute{
Philosophy Department\\
Watson Hall\\
Queen's University\\
Kingston ON, K7L 3N6, Canada\\
\and \\
School of Computing\\
557 Goodwin Hall\\
Queen's University\\
Kingston ON K7L 2N8, Canada\\
\email{c.stinson@queensu.ca}
}
\date{January 2022}

\maketitle

\begin{abstract}
Algorithmic bias has been the subject of much recent controversy. To clarify what is at stake and to make progress resolving the controversy, a better understanding of the concepts involved would be helpful. The discussion here focuses on the disputed claim that algorithms themselves cannot be biased. To clarify this claim we need to know what kind of thing `algorithms themselves' are, and to disambiguate the several meanings of `bias' at play. This further involves showing how bias of moral import can result from statistical biases, and drawing connections to previous conceptual work about political artifacts and oppressive things. Data bias has been identified in domains like hiring, policing and medicine. Examples where algorithms themselves have been pinpointed as the locus of bias include recommender systems that influence media consumption, academic search engines that influence citation patterns, and the 2020 UK algorithmically-moderated A-level grades. Recognition that algorithms are a kind of thing that can be biased is key to making decisions about responsibility for harm, and preventing algorithmically mediated discrimination.

\end{abstract}


\section{Introduction} \label{intro}
Artificial intelligence (AI) has been under fire after the revelation of a series of cases where algorithms meant to help make decisions like predicting recidivism in court cases \citep{Angwin2016}, hiring the most qualified employees \citep{Raub2018}, or recognizing people's faces \citep{Buolamwini2018} have turned out to discriminate along racial and/or gender lines.\footnote{Additional fuel on this fire includes rampant workplace sexual harassment (\url{https:// money.cnn.com/technology/sexual-harassment-tech}), complicity in the erosion of democracy (\url{https://www.democracynow.org/2020/1/10/2020_election_digital_manipulation_cambridge_analytica}), flouting of international law (\url{https://www.huffingtonpost.ca/2019/04/25/canada-s-privacy-czar-denounces-facebook-says-it-won-t-admit-it-broke-law_a_23717150}), and hiring PR firms to discredit legal scholars who advocate for regulation (\url{https://www.cnet.com/features/uber-lyfts-fight-over-gig-worker-status-as-campaign-against-labor-activists-mounts}).} A recent high-profile scandal involved the UK's use of an algorithm to assign grades to students in place of the university entry exams that were cancelled because of COVID-19. The grades of students at state schools tended to be lowered compared to teacher assigned grades, whereas students at private schools were more likely to see their grades increase, resulting in days of angry protest against class discrimination, where students shouted slogans like ``Fuck the algorithm''\footnote{\url{https://novaramedia.com/2020/08/17/fuck-the-algorithm-how-a-level-students-have-shown-future-of-protest}}.

This general problem of algorithms producing discriminatory results is referred to as \emph{algorithmic bias}, yet the role of algorithms themselves in bringing about these biased outcomes is a matter of some controversy. 
The goal here is to provide an analysis of the question, ``can algorithms themselves be biased?'' by clarifying the terms involved, then using this analysis to argue that algorithms themselves can be biased. Questions about moral or legal responsibility for bias are distinguished from questions about the location of bias, several relevant meanings of `algorithm' and `bias' are disambiguated, then a few cases are examined in some detail. The result is that there are multiple locations or stages where bias can take root, including not just data gathering and management, but also algorithm design and model building.

These clarifications are motivated in several ways. One is that discussions of algorithmic bias among AI researchers and scholars who study AI as a social phenomenon can be rather acrimonious, involving cycles of misunderstanding, equivocation, and talking past one another \footnote{\url{https://venturebeat.com/2020/06/26/ai-weekly-a-deep-learning-pioneers-teachable-moment-on-ai-bias/}}\footnote{\url{https://syncedreview.com/2020/06/30/yann-lecun-quits-twitter-amid-acrimonious-exchanges-on-ai-bias/}}. Another motivation, which partly explains the tone of those conversations, is that having a reputation for racial and/or gender discrimination is damaging to the reputation of AI in general, machine learning (ML) in particular, and the major tech companies that are leaders in ML research. ML researchers have an interest in fixing the problem, or at least muting the conversation. Another important motivation is the moral imperative to identify the causes of discrimination so that it can be effectively prevented, and appropriate reparations can be made.

The moral status of algorithms has been explored extensively within social science and media studies, often from historical, critical theory, or social justice perspectives. One notable example that touches on the more conceptual themes that are the focus here is \cite{FriedmanNissenbaum1996} who explore the idea that computing systems can be biased. 
However, algorithmic bias remains a relatively untouched topic within analytic philosophy. \cite{Johnson2020} explores connections between algorithmic bias, implicit bias, and stereotypes, but considers only biased datasets, not bias in algorithms themselves, so sidesteps the main question at stake here. \cite{Hancox-Li2021}, which discusses the values embedded in feature importance methods from the perspective of feminist epistemology, is the only analytic philosophy paper we are aware of that addresses the value-ladenness of the decisions algorithm designers make.

Section 2 breaks down what `algorithm' and `bias' mean to sort out the trivial from the interesting interpretations of the claim that an algorithm is biased. Section 3 explores some statistical biases that are well known among ML researchers to affect algorithms. Section 4 connects these statistical biases to bias of moral import, explores how an algorithm that is suitable for some tasks and datasets might be biased in other contexts, and connects to the more general question of whether algorithms can be political \citep{Winner1980} or oppressive \citep{Liao2020}. Section 5 offers concluding reflections on why it is important not to be misled by the neutral appearance of algorithms.


\section{On `algorithm' and `bias'}\label{sec:1}
The lines of disagreement over algorithmic bias were put on full display in a Twitter battle that erupted in June 2020 between inventor of deep learning and Chief AI Scientist at Facebook, Yann LeCun, and Timnit Gebru, an AI researcher who founded the Black in AI affinity group, and co-led Google's AI Ethics team until she was ``resignated''\footnote{As discussed in this \href{https://venturebeat.com/2020/12/10/timnit-gebru-googles-dehumanizing-memo-paints-me-as-an-angry-black-woman/}{interview}, the term ``resignated'' has been used by Gebru and her former colleagues to describe  her controversial departure from Google.} in December 2020. An image was circulating where an upsampling tool that takes blurry pictures and makes them sharp had upsampled a fuzzy but easily recognizable Barack Obama into a blue eyed white man (see Figure 1). The image was accompanied by comments about the dangers of bias in AI. LeCun quickly replied, ``ML systems are biased when data is biased. This face upsampling system makes everyone look white because the network was pretrained on FlickFaceHQ, which mainly contains white people pics. Train the *exact* same system on a dataset from Senegal, and everyone will look African.''\footnote{\url{https://twitter.com/ylecun/status/1274782757907030016}}. 

\begin{figure}[t]
\centering
\includegraphics[width=0.7\columnwidth]{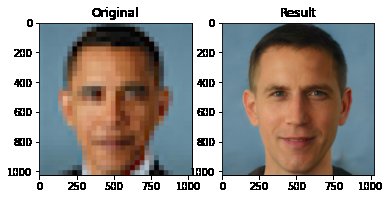}
\caption{Face upsampling on a picture of Barack Obama. Tweeted by @Chicken3gg \url{https://twitter.com/Chicken3gg/status/1274314622447820801}}
\label{fig1}
\end{figure}

This suggestion that algorithmic bias is caused by bias in the dataset used to train the model, and that de-biasing the dataset is the appropriate fix is a point LeCun and other prominent AI researchers have made several times before, as in a 2019 tweet by LeCun, shown in Figure 2, which prefaced a link to a December 6th New York Times article arguing that biased data is easier to fix than biased people. 

\begin{figure}[t]
\centering
\includegraphics[width=0.6\columnwidth]{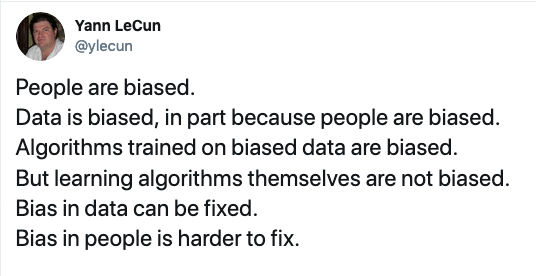}
\caption{Screenshot of 2019 Tweet by Yann LeCun, taken by the author
}
\label{fig2}
\end{figure}

LeCun's tweets express a popular view among AI researchers, that blame for algorithmic bias should be assigned to either people or data (but preferably data). The suggestion that algorithms themselves might also be among the causes of algorithmic bias is swiftly swept aside. A very common reaction is to treat the claim that algorithms can be biased as a category mistake. Algorithms are math, and math is objective, therefore algorithms are morally neutral, according to reply guys. Anyone who dares to point out the fallacies in that argument is ridiculed as a wokescold who doesn't understand math. For instance, a recent headline reads, ``It's not just a joke anymore: They're actually claiming math is racist''\footnote{\url{https://www.washingtonexaminer.com/its-not-just-a-joke-anymore-theyre-actually-claiming-math-is-racist}}. There are, however, reasons to think that algorithms themselves are also part of the problem. 

\subsection{What is an algorithm?}\label{sec:2.1}
The first step in deciding whether algorithms can be biased is to clarify what that claim could mean. One site of confusion and equivocation is ambiguity in the meaning of `algorithm'. When `algorithm' is understood as just bits of math and logic, it is easy to ridicule the idea that algorithms are biased. Exclusive OR is not elitist, despite the name, and the number 55378008 is not complicit in rape culture, despite math class sexual harassers' penchant for showing it upside down on calculators. 

In critical discussions of algorithmic bias, `algorithm' typically has a broader scope, referring to computational systems embedded in social contexts. The Algorithmic Accountability Act introduced to the US Congress in 2019 (H.R.2231, \citeyear{H.R.2231}) refers to algorithms in the title, but the text of the bill is about automated decision systems, like the use of facial recognition systems by police forces, or automated hiring systems by corporations. It is just as obvious that these automated decision systems can be biased as it is that ``$2 + 2 = 4$'' is not\footnote{But see the discussion following this Twitter thread by Kareem Carr \url{https: //twitter.com/kareem_carr/status/1289724475609501697} defending the claim that sometimes 2+2=5.}. That people mean different things by `algorithm' is one reason for the failure in communication. 

Here `algorithm' will be used in an intermediate way, to refer to more than just a set of mathematical or logical operations, but less than an entire computational system embedded in a social context.\footnote{Except in the phrase `algorithmic bias' which I take to refer broadly to bias in computational systems.} This intermediate scope reflects typical uses of `algorithm' in computer science, as is reflected in the IEEE Standards definition, ``A finite set of well-defined rules for the solution of a problem in a finite number of steps... Any sequence of operations for performing a specific task'' \citep{IEEEGlossary}. 
An algorithm is a set of rules or operations, not an entire computational system, but a set of rules \emph{for a specific problem or task}. The same set of rules might be unobjectionable for one task, but unfair or discriminatory for a different one.

Alphabetical order, for example, taken as a rule in abstracta may not be the sort of thing that could be biased. Alphabetical order used to organize people into groups that will be treated equally (like the lines at the registration desk at a conference) is fair. But alphabetical order used for the distribution of scarce resources would systematically benefit the Aaron Abrahams of the world at the expense of the Zara Zhangs. In some contexts the use of alphabetical order could even constitute racial discrimination, given that name frequency among ethnic groups are not evenly distributed across the alphabet, as demonstrated in Figure 3. First letter of last name can act as a proxy for race. 

\begin{figure}[t]
\centering
\includegraphics[width=.8\columnwidth]{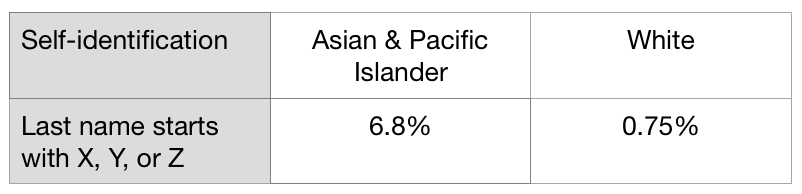}
\caption{Prevalence of last names near the end of the alphabet, by race, calculated by the author based on 2000 US census data available at www.namecensus.com. An order of magnitude more people with last names starting with X, Y or Z identified as ``Asian \& Pacific Islander'' than ``White''. Other racial groups had prevalence similar to ``White''.}
\label{fig 3}
\end{figure}

A scenario where use of alphabetical order arguably led to racial discrimination played out recently. Toronto area school boards used alphabetical order to assign children attending online school in 2020/2021 to virtual classrooms. There were, however, delays in hiring enough teachers for those virtual classrooms. An unforeseen (but perhaps predictable) consequence was that students with last names starting with Z, the overwhelming majority of whom are Chinese-Canadian, had to wait an extra two weeks before their classes could start, because the classes at the end of the alphabet hadn't yet been assigned teachers.\footnote{\url{https://www.theglobeandmail.com/canada/article-toronto-area-school-board-sorts-online-classes-alphabetically-raising/}} 

\citet[332]{FriedmanNissenbaum1996} write, ``A system discriminates unfairly if it denies an opportunity or a good or if it assigns an undesirable outcome to an individual or group of individuals on grounds that are unreasonable or inappropriate.'' In this case, a good (two weeks of schooling) was unfairly denied to people based on a proxy for race. This example demonstrates how given real-world distributions of human characteristics, an apparently neutral procedure can cause discriminatory outcomes \emph{for some tasks}, even though it might seem ridiculous to call alphabetical order racist. The task is critical. On another occasion, I was proctoring an exam in a hall where the last alphabetical section of a huge undergraduate math course at [anonymized university in large North American city] 
was writing. All but one of those approximately 80 students whose last names began with Z were Asian. However, the exam hall had decent lighting, extra paper, and sturdy desks just like all the other exam halls where the rest of the class was writing, so while it was a case of accidental racial sorting, nobody was being denied an opportunity or a good. 

This simple example illustrates a pattern that plays out in many other contexts. This example also demonstrates a flaw in LeCun's proposal that the solution is to fix the dataset. Here, fixing the dataset would mean forcing new last names on people. Some comments on news articles about the issue in fact suggested this fix: ``Maybe some cultures need to spice up their naming choices''\footnote{ibid.} said one commenter. This would, however, constitute a human rights violation (and was presumably meant as a joke). Real world data about people is very often unevenly distributed, so this scenario is in no way exceptional. In some cases the lack of uniformity can be patched by manipulating datasets to be more inclusive of the groups who would otherwise be underserved, but not always.  

\subsection{What is bias, and where does it reside?}\label{sec:2.2}
`Bias' is still more ambiguous than `algorithm'. Friedman and Nissenbaum's groundbreaking (\citeyear{FriedmanNissenbaum1996}) paper, ``Bias in Computer Systems'',  distinguishes value neutral notions of bias, such as how it is used in Statistics and ML as a technical term with no moral connotations from ``bias of moral import''. Bias of moral import is treated as synonymous with unfair discrimination. They divide bias of moral import into three types, based on an analysis of the source of the bias: preexisting bias, technical bias, and emergent bias. 

As the name suggests, preexisting bias is at least in part distinguished temporally. These are biases that ``exist independently, and usually prior to the creation of the system'' \citep[334]{FriedmanNissenbaum1996} either in the society at large, or in individuals involved in the design of the software, and include both conscious and unconscious or implicit biases. Emergent bias also has a partly temporal character. It arises when users interact with the system, and ``typically emerges some time after a design is completed'' \citep[335]{FriedmanNissenbaum1996}. Examples include cases where new knowledge can't be incorporated into the design after the fact, and where users have different abilities, needs, or values than were anticipated by designers, such as giving written instructions to an illiterate population. Technical bias is more of a hodge-podge of examples, including screen size limiting visibility of options, misuse of pseudo-random number generators, imperfect formalization of ambiguous or complex concepts, or algorithms used in contexts for which they are inappropriate.

Friedman and Nissenbaum's taxonomy of bias is a good starting point, but is not fine-grained enough for contemporary discussions of algorithmic bias. Consider the case of Street Bump, a smartphone app that detects bumps during car rides, then reports the location of the bumps to a central system for allocating road repair resources. As \cite{Crawford2013} points out, the app fails in its goal to make allocation of road repair resources fairer, because smartphones are much more likely to be found in wealthier neighborhoods than poorer ones. Is this preexisting bias, because the unequal distribution of technological resources exists independently in society at large? Is it technological bias, because the app depends on technologies only found in certain types of phones? Or is it emergent bias, because there was a mismatch between the actual abilities of users and the abilities anticipated by the designers? It is hard to say.

The contemporary taxonomy that distinguishes between biased data, biased people, and biased algorithms (if those exist) is also not fine-grained enough. The first problem with this taxonomy is that it conflates two different kinds of causes. The question of who, if anyone, bears moral or legal responsibility for algorithmic bias is orthogonal to the question of where in the computational system the bias occurs. Greater clarity can be found by distinguishing bias along these two axes: responsibility and location. Knowing that the dataset is the location of the bias does not tell you much about whether the people who gathered the data were acting irresponsibly, and knowing that the coders had pure intentions does not tell you much about whether the model is biased. 

The people responsible for a biased system could be intentionally and knowingly discriminating, as in attempts to Gerrymander election districts so as to disenfranchise some voters, or accidentally creating a discriminatory system, as in Street Bump and the Toronto school board example. Heather Douglas helpfully distinguishes the moral behaviour of scientists into cases of negligence, recklessness, and responsibility. Knowingly creating an unreasonable risk is recklessness. Negligence is when one is not aware of risks, but should be \citep[61]{Douglas2003}. Many cases of algorithmic bias involve decision-makers failing to anticipate harms, but whether they should have been aware of those harms is sometimes hard to say. The Toronto School Board case is one that might fall in this grey area.

A clearer case of negligence is a much maligned ML paper by Wu and Zhang that claims to be able predict ``criminality'' from analysis of people's faces \citep{Wu2016}. This study re-invents Francis Galton's 19th century eugenics techniques with contemporary ML tools. ID photos of people convicted of crimes and people presumed not to be criminals are used as data to train a ML classifier using biometric facial recognition tools to predict ``criminality'' from the shape of people's faces. While the study claims near perfect performance at detecting faces associated with convicts, it also mis-labels non-convicts as convicts more than half the time, making it useless even for its highly questionable purpose.\footnote{\url{https://aeon.co/ideas/algorithms-associating-appearance- and-criminality-have-a-dark-past}} As there is no shortage of information available explaining why Galton's eugenics program was both morally objectionable, and scientifically suspect, as well as plenty of evidence to suggest that criminal justice systems often make discriminatory judgments about which people to treat as suspects, which people to charge, and which people to convict, it is reasonable to say that the authors of this study ought to have have realized that the study could be harmful. While Wu and Zhang apparently did not anticipate any harm or see the backlash coming, it is fair to call this a case of negligence.

In cases of discrimination rooted in systemic social systems, such as when the lack of minority representation in a field influences which research questions are pursued, and how data is collected \citep{West2019a}, discrimination may be non-accidental, but also not the fault of individual researchers. 
That facial recognition algorithms are an order of magnitude less accurate at determining gender for Black female faces than for white male faces \citep{Buolamwini2018} is attributed to the lack of Black and female faces among the training data used to build facial recognition systems. That the datasets commonly used in vision science lack gender and racial diversity is due to systemic factors like lack of diversity in computer science departments, and geographic skew in research opportunities. 

Although much more could be said about the responsibility axis (see \cite{Barocas2016}), I will focus now on the location axis. Datasets are not the only location where bias can occur; there are many stages in the workflow of a ML system, and bias can creep in at any one of those stages, including: problem selection, data collection, algorithm design, model building, and use. A simplified workflow of these stages is shown in Figure 4. Additional stages like encoding of data, calibration, testing, and re-design might also be included. 
\begin{figure}[t]
\centering
\includegraphics[width=\columnwidth]{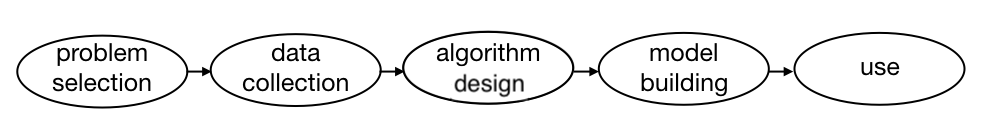} 
\caption{Simplified workflow of a ML system. Bias can affect any of these stages.}
\label{fig4}
\end{figure}

An example of biased problem selection is the attempt to predict criminality from people's faces. Looking for correlations between criminal convictions and facial features to support tools for predicting criminality, despite historical and sociological evidence suggesting that these correlations result from bias in the criminal justice system, skews toward the perspective of police, and the status quo they protect. Databases used to train facial recognition systems that consist of mostly white male faces are an example of biased data collection. An example of bias in use is how ML-powered surveillance technologies like Clearview AI and StingRay are available to many police forces,\footnote{\url{https://theintercept.com/2020/07/14/microsoft-police-state-mass-surveillance-facial-recognition/}} but not to policed communities, unless they build their own.\footnote{\url{https://www.nytimes.com/2020/10/21/technology/facial-recognition-police.html}}

The two locations in this simplified workflow that plausibly count as where ``algorithms themselves'' reside are algorithm design and model building. An example of biased algorithm design is how YouTube's recommendation system was optimized for engagement (i.e., watch time, clicks) rather than other possible metrics, such as user satisfaction or reputability of information sources \footnote{\url{https://www.theguardian.com/ technology/2018/feb/02/how-youtubes-algorithm-distorts-truth}}. This puts the needs of stockholders and CEOs, who have an interest in maximizing ad revenues, above the needs of users and broader society, who might value things like curbing misinformation, and preventing internet addiction. 

Optimizing for engagement has been blamed for YouTube's tendency to serve viewers ever more radical content, sowing political division. The Christchurch shooter described YouTube as a ``significant source of information and inspiration'' \citep[4.4.46]{NZRC2020}. There is no evidence that radicalization was anticipated by the team developing YouTube's algorithm as an outcome of optimizing for engagement.\footnote{There is evidence of a delay in acting to prevent these outcomes after it was known, however.\url{https://www.wsj.com/articles/
how-youtube-drives-viewers-to-the-internets-darkest-corners-1518020478}} Before these disturbing trends came to light, engagement might have seemed like a reasonable proxy for user satisfaction. That YouTube hosts vast troves of disreputable, obscene, and disturbing content, and that it is primarily used for entertainment makes engagement a poor choice of metric in retrospect, but for a different task, optimizing for engagement might be appropriate.

The UK A-levels controversy provides an example of bias in model building. In a typical year, students are given provisional university places based on mock exams, which their teachers grade, but those places can be rescinded if the student does not achieve a high enough grade on a final standardized exam. Usually, mock exam grades are slightly higher than final grades. In 2020, COVID-19 led to the cancellation of the final standardized exam. Instead of accepting the teachers' grades as final, an algorithm was used to adjust students' scores, with the desired overall effect of adjusting scores downward to put them in line with previous years' exam scores. However, the uneven distribution of those downgrades, and some anomalous individual cases led to mass protests. Students from fee-paying private schools on average ended up with grade increases, while students from state schools in poor neighborhoods ended up with decreased grades on average. In some cases students with A* mock exam grades were demoted to Cs, and lost their places at elite universities. 

Although multiple issues have been uncovered with the design and implementation of the model used to predict exam grades\footnote{See \url{https://thaines.com/post/alevels2020} for analysis.}, perhaps the biggest issue identified is that the historical performance of schools weighed too heavily into what grade an individual student could be assigned. If your school had never before graduated an A* student, the model could not predict an A*, regardless of your individual record. If your school had a record of failing x\% of students, and you were in the bottom x\%, the model would predict a failing grade, regardless of your individual record. The problem here is not bias in the training data. Every school's historical record of exam results was included, and despite being imperfect measures of educational achievement, standardized exam results are arguably less biased than teachers' grades. The problem was also not with the algorithm design choice to build a model that predicts exam results based on past relationships between grades and exam performance at particular schools. It is reasonable to make adjustments in line with the historical relationship between teacher grades and exam results at a given school, assuming teacher turnover is low, and school grading culture is relatively stable. The problem lies in the particular way that the model learned to predict exam results for individuals. Constraining the grades that could be predicted to follow the pattern of a school's historical record in such detail is an unfair way of assigning a score that is meant to reflect individual achievement.\footnote{A wide variety of things get called a ``model'' within ML, and ``model'' is sometimes used interchangeably with ``algorithm''. I have tried to choose an example where ``model'' being the correct word should be fairly uncontroversial, but one could quibble.}

\section{Value neutral bias}
When dealing with messy real-world examples, it is difficult to separate the responsibility axis from the location axis. We now focus on the value neutral branch of Friedman and Nissenbaum's taxonomy of bias, covering statistical bias, and in-principle demonstrations of bias in algorithms in order to put the focus more squarely on the location axis.

\subsection{Statistical bias}
Bias is widely used as a neutral technical term in ML, to refer to a collection of statistical properties of data, data collection, and data measures. Two of the many kinds of bias relevant to ML are selection bias, and estimator bias. 
\emph{Selection bias} is when a non-random process is used to select a sample from a population, such that members of the population do not have equal chances of being selected. For example, if people conducting a survey preferentially recruit tall people to participate, perhaps by posting the signup sheet near the top of a bulletin board, the resulting sample would be biased, and the survey results might be misleading. Selection bias in data collection tends to lead to worse performance of algorithms trained on those data. 

Several prominent examples where datasets used in AI are said to be biased involve selection bias. 
That the collection of faces on which facial recognition software is trained contains a far greater proportion of pictures of white men than their representation in the populations where that software is used is the result of a selection bias. Faces were not drawn from the population at random. To fix a selection bias in a dataset, examples from the sectors of the population not represented would need to be added. Some cases of bias in policing applications involve a more subtle selection bias, where the dataset may contain everyone convicted of a crime (no selection bias here), but because police and judges make non-random decisions about who they stop and search, who they suspect of crimes, which crimes they deem worthy of attention, and whose testimony they believe, the people convicted of crimes are not selected at random from the population of people who commit crimes. The quality of that non-randomness involves bias of moral import, but we can nevertheless keep the statistical notion distinct. If selection bias in data collection is the only bias present, then the strategy of adding more examples from the excluded categories to the data would be an appropriate solution. 

\emph{Estimator bias} is the extent to which the value of a variable measured in a sample differs from the value of that variable measured in the whole population. For example, if the average height among a sample of people was 210cm, whereas the average height for the population was 170cm, that discrepancy is the estimator bias. Estimator bias can occur as a result of selection bias (such as in the example of a survey signup sheet posted too high), but it can also occur in a sample chosen at random. As \cite{Zhao2020} discusses, random sampling typically results in unbalanced samples.

A related statistic is \emph{variance}, which indicates how far a set of measures are spread out from their average value. For example, the average height in a sample could be identical to the population average of 170cm, either because the varied heights of the people in the sample averaged to 170cm, or because everyone in the sample happened to be 170cm tall. A mismatch between a sample and a population's variance is another way that a sample can be unrepresentative. In the next section we will see examples where focusing only on the average performance of an algorithm, and ignoring the variance in an algorithm's performance across the population can lead to discrimination.  


\subsection{Statistical bias in ML applications}\label{sec:3}
One area of ML where statistical biases have been well documented is information retrieval, which includes tasks like recommendation and search. 
A recommendation service presents its users with items (songs, movies, books, etc.) that it predicts the user will like. Typical ways of measuring user preferences are star systems, ``likes'', clicks, purchases, views, and listens. There are, however, a number of documented biases in recommendation, meaning that what is recommended to users differs systematically from their true preferences. Recommendations are often inaccurate (estimator bias) or too narrow (too little variance). Often the cause of the problem is a selection bias.

The simplest of these biases is dubbed the ``cold-start problem'. The cold-start problem is worst when collaborative filtering algorithms are used to make recommendations. Collaborative filtering bases recommendations on what similar users prefer, rather than on the content of the item. This means that newly released items will not be recommended at all, since they have not yet been seen by any users, hence are not known to be liked by anyone. This type of algorithm creates a selection bias in ratings across items. A related issue is ``popularity bias'' \citep{Herlocker2004} where very popular items are over-recommended, relative to their utility as recommendations. As an item becomes more popular, there will be more ratings in the system for that item, introducing another selection bias. Both of these biases can have a homogenizing effect on recommendations, increasing the visibility of already popular items, and steering users toward them. Cold-start and popularity bias were unintended side-effects of the design of collaborative filtering algorithms, not desired features. See \cite{Stinson2022} for more details on bias in recommendation.

Similar statistical biases affect search engines. The equivalent of popularity bias is called the ``winner-take-all'' effect in search \citep{Goldman2008}, where top placement in search begets more top placements, because users are most likely to click on the top search results. A similar problem where some websites (particularly ones outside the US) are not ranked as hits despite relevance to search terms, is termed ``coverage bias'' \citep{Vaughan2004}. ML researchers working on search clearly state that there is ``bias that arises from the data that serves as the input to the ranking system and the bias that arises from the ranking system itself'' \citep{Kulshrestha2017}. In other words, both biased data and biased algorithms are well known to exist in ML. It is utterly uncontroversial among ML researchers that information retrieval algorithms manifest a host of statistical biases.  

Illustrations of these biases include the many complaints that can be found online about culturally inappropriate recommendations, like white hairdressers being first in the rankings for search terms like `Black', `relaxer', and `natural', or Christmas movies being aggressively recommended every November to non-Christians. Another type of complaint arises when the recommender system overfits to an essentialized version of a minority identity. For example, after viewing a season of \emph{Rupaul's Drag Race} your Netflix recommendations might become overwhelmed with LGBTQ content otherwise unrelated to your viewing habits. Even when the recipient of those stereotyped recommendations do happen to like some of what is recommended on that basis, a wrong is still being committed. As \cite{Basu2019} argues, racist (and by the same logic homophobic) beliefs that happen to be correct can nevertheless be harmful. 
These biases are also evident in academic research practices. A recent study \citep{West2019} suggests that GoogleScholar's search results may have had a homogenizing effect on citation practices. More citations are going to the top 5\% of papers by citation count, and a smaller proportion of papers are being cited overall since the release of GoogleScholar. 

There are well-known phenomena that constitute bias in algorithms themselves outside of information retrieval too, though these are usually known by a different name. \cite{Hooker2021} points out that if you call it `test-set accuracy' rather than `algorithmic bias' most people in ML would be familiar with the point that ``modeling choices---architecture, loss function, optimizer, hyper-parameters---express a preference for final model behavior.'' The design of an algorithm affects how it will work, including the effects of different segments of training data, and which performance metrics are prioritized. \cite{Hancox-Li2021} elaborate on how feature importance methods, which are used to generate explanations of algorithmic decisions, make unjustified claims about which features are desirable in an explanation that may conflict with the equity goals of explainable AI.

Even with synthetic data (machine generated datasets with pre-set statistical properties), standard ML algorithms can be shown in principle to be biased. For example, the k-means algorithm is one of ML's most basic clustering tools, which partitions a set of data points into k clusters, such that the sum of the distances of each point to it's cluster centre is minimized. \cite{Ghadiri2021} illustrate the unfairness of k-means with an example where the dataset consists of a larger group that is fairly spread out, and a smaller group at a distance. Because the objective function is to minimize the total distance of points to centres, the solution for k=2 will be to place both cluster centres within the larger group. All of the smaller group's points will be far from the cluster centres, and essentially treated as outliers. With a simple alteration to the objective function, \cite{Ghadiri2021} create a fair k-means algorithm that assigns one cluster to each of the two groups.

\section{From Value Neutral Bias to Discrimination}\label{sec:4}

This brings us to how statistical bias in ML can translate into bias of moral import. The fact that inequality often falls out as a result of using algorithmic decision-making systems is a point that has been made many times by historians, social scientists and media theorists. Two notable examples are \cite{Noble2018} who documents the ways that search algorithms fail to serve the needs of Black women, and \cite{Eubanks2018} who documents how algorithmic decision-making in social services exacerbate class inequities. The missing piece for our purposes is to detail how statistical bias in algorithms themselves connects to discriminatory outcomes.

There is empirical evidence suggesting that statistical bias in search and recommendation can produce unfair outcomes. Examples include media producers not all having a fair chance of their products being seen \citep{Mehrotra2018}, searches for political information on Twitter being biased toward the most popular posts, potentially influencing political views \citep{Kulshrestha2017}, and differences in the utility of recommendation systems for different demographic groups \cite{Ekstrand2018, Neophytou2021}. There are also demonstrations that recommendation algorithms can produce results that are more biased than the data used to train them \citep{Ekstrand2018a,Tsintzou2018}.
The implications go well beyond unwanted recommendations. As algorithms mediate more and more of our access to information, access to services, and decisions about our lives, their performance becomes a significant equity issue. 

Where ML is used in critical, life-altering decisions, choices of architecture, loss function, optimizer, or hyper-parameters can mean that the model will learn more from examples belonging to majority groups than minority groups, or prioritize getting the right answer for majority groups at the expense of minority groups. Either scenario can lead to a final product that works badly for some groups of people. Two examples of ML being used in real world applications despite potentially dangerous consequences are blood oxygen monitoring tools that were calibrated for lighter skin, so are less accurate for people with dark skin, often failing to detect clinically important oxygen levels \citep{Feiner2007}, and differential privacy methods for anonymizing sensitive data that put outliers in data, like people with disabilities, at risk of either being easily re-identified, or having their data blurred to such an extent that they are no longer represented \citep{Stinson2018}.

Several of the biases described above stem from algorithm designers choosing to maximize the mean accuracy of classifications or predictions as their objective function. The effect is a tendency to zero in on what works best for the majority. This ignores not only the value of ``information diversity'' \citep{Bozdag2013}, but also disproportionally disadvantages minority users. In many cases, members of minority groups are literally on the margins of distributions of human traits \citep{Treviranus2014}, while members of majority groups tend to be found nearer the middle of the distribution. Designing technologies to work well for the majority clustered around the mean not only disadvantages people who occupy the tails of the distribution, it wastes an opportunity to make use of the knowledge available on the margins.\footnote{\url{https://medium.com/@jutta.trevira/inclusive-design-the-bell-curve-the-starburst-and-the-virtuous-tornado-6094f797b1bf}} 

Maximizing mean accuracy is often taken for granted as an obvious goal one should have in evaluating the performance of a ML model, but it is a choice that researchers make. It is a choice that makes sense when the data is normally distributed, but it is less justifiable when the data is is multi-modal or has long tails. As \cite{Hancox-Li2021} argue, this practice of treating some properties as universal success criteria for ML systems ``ignores the fact that those properties may not be the most appropriate ones in every context of application,'' and a better strategy would be to design the objective function around ``what users need in specific contexts'' \citep{Hancox-Li2021}. For medical applications like blood oxygen monitoring, minimizing false negatives should have been among the objectives.

\cite{FriedmanNissenbaum1996} define bias in terms of outcomes; a technology that unfairly denies goods to some groups is discriminatory, regardless of the intentions of the technology's designers, or how that outcome comes about. The examples discussed above are ones where AI applications are biased in this sense of having outcomes that are slanted in favour of some groups of people at the expense of others. In many of the examples discussed the slant is to the disadvantage of protected groups, which might constitute illegal discrimination. Bias can also imply a deeper moral issue than just a slanted outcome; it can mean a political motivation, or a system of oppression at play.

Earlier the responsibility axis was distinguished from the location of bias so as not to confuse algorithm designers and social forces with specific locations of bias, obscuring the existence of bias in algorithms themselves. In order to inquire whether algorithms are biased in one of the deeper senses, we now need to bring people and social forces back into the picture. Whether the people who choose, for example, to maximize mean accuracy, belong to a group with political motivations or social power, and are acting in response to their social position (whether unintentionally, negligently, or recklessly) may not matter when deciding on the location of bias, but it does matter when deciding whether that bias has a political tenor.

Langdon Winner famously argued that artifacts can have politics in two ways: when technology is designed to settle a political issue, or when a technology ``appear[s] to require, or to be strongly compatible with, particular kinds of political relationships'' \citep{Winner1980}. Maximizing mean accuracy is not a neutral choice of objective function. In Winner's sense, it is the second type of political artifact, in that it is strongly compatible with prioritizing the needs of the majority.\footnote{I do not mean to imply that maximizing mean accuracy is a near universal property of ML systems, nor that it is the only such algorithmic design choice that could be described as political in this sense. It is intended here as an illustrative example.}

\cite{Liao2020} extend Winner's analysis into a discussion of oppression. ``Oppressive things'' need to meet three conditions: being biased in the same direction as an oppressive system, being the causal product of the oppressive system, and in turn exerting psychological or social influence that supports the oppressive system \citep{Liao2020}. 
Take the example of a recommender system that is statistically biased toward recommending more popular items than its users want, with the result that works made by and for minorities are under-recommended compared to both the proportion of positive ratings of those items in the training data, and in terms of user satisfaction measures. Suppose that the choice of objective function was influenced by the fact that the development team consisted of mostly middle-class white males with mainstream tastes, and that it did not occur to anyone on the development team that their algorithm might not work well for people on the margins, because they have never personally encountered a problem like their one-size-fits-all T-shirt not fitting, their crayon box not including a colour that matches their skin, an exam being scheduled on their religion's most important holiday, or films in their first language being classified as `foreign'. As a result of the recommendation system under-recommending works by and for minorities, even the users who would choose those works fail to find them, keeping the ratings low, and the authors' attempts to fund further works fail because of low sales.
In this case this bias is in the same direction as oppressive systems in society at large, the reason why the recommender system was designed this way was because of systemic bias's effects on the composition of the ML workforce, and the outcome further supports that bias by artificially reducing the market for media by and for minorities, and psychologically harming people unable to find media that represents them.

Another example that is clearly congruent with the direction of oppression is how automatic soap dispensers fail to work well for people with dark skin \citep{Fussell2017} because their operation depends on light bouncing back from the user's hands. This likely was caused by technology designers de-prioritizing the needs of dark skinned users. It likely also exerts a psychological influence that supports racism. The UK A-levels case is clear on criteria 1 and 3. The bias disadvantages lower class students, and entrenches their class disadvantage by denying them educational opportunities. It seems likely that the design choices that led to the mess were also caused by that class system in that the designers failed to consider the possibility that a talented student might show up at a historically underperforming school. The algorithm embedded the assumption that people from lower class neighbourhoods will remain lower class. These examples qualify not just as biased algorithms, but as oppressive algorithms.

\section{Conclusions}\label{sec:5}

False claims about the neutrality of algorithms discourage further research into discovering and fixing bias in algorithms. Perhaps the greatest danger posed by claims that algorithms themselves cannot be biased is that the illusion of neutrality can be exploited.

The UK A-levels case is one where an attempt was made to pass off a biased algorithm as neutral, on the basis of it being mathematical in nature and somewhat complex. Another case saw the trustees of the Stanford Hospital pass the buck to ``the algorithm''  for leaving residents who work directly with COVID-19 patients out of the first round of vaccinations, while more senior staff with no direct patient contact made the list \footnote{\url{https://www.technologyreview.com/2020/ 12/21/1015303/stanford-vaccine-algorithm/}}. These are just the cases that we know about, because they made the news, garnering enough attention to question and reverse the unfair algorithmic decisions. There must be many more cases we do not know about.

A non-trivial way of understanding, `Can an algorithm be biased?' is as a question about whether the design choices made about algorithms can lead to bias of moral import, independently of any bias that may affect other stages of the ML workflow, like data collection. It is abundantly clear that statistical bias affects many ML algorithms. There is also considerable evidence suggesting that these statistical biases can lead to discrimination. In some cases they fit the criteria for being called oppressive. Whether the choice to use a biased algorithm is to be blamed on a biased person or institution, or considered accidental is an orthogonal question. Algorithms themselves can be biased in all relevant senses. Fixing biased datasets and improving the ethical behaviour of AI workers are absolutely necessary steps, but they will not eliminate all bias in ML. The claim that algorithms are neutral is both false and dangerous. 


\bibliographystyle{spbasic}
\bibliography{ML}{}

\begin{thebibliography}{38}
\bibitem[{Stinson(forthcoming)}]{Stinson2022}
Stinson C (forthcoming) Algorithms are not neutral: Bias in collaborative
  filtering. AI and Ethics
\end{thebibliography}

\end{document}